\documentclass[manuscript]{aastex}

\shorttitle{Galactic Planetary nebulae}
\shortauthors{Stanghellini \& Haywood}

\begin{document}

\title{The Galactic structure and chemical evolution traced by the population of planetary nebulae}

\author{Letizia Stanghellini}
\affil{National Optical Astronomy Observatory, Tucson, AZ 85719}
\email{lstanghellini@noao.edu}

\and

\author{Misha Haywood}
\affil{GEPI, Observatoire de Paris, CNRS, Universit\'e Paris Diderot, 92190 Meudon, France}
\email{Misha.Haywood@obspm.fr}

\begin{abstract}

Planetary nebulae (PNe) derive from the evolution of $\sim$1--8 M$_{\odot}$ mass stars, corresponding to a wide range of progenitor ages, 
thus are essential probes of the chemical 
evolution of galaxies,  and indispensable  to constrain the results from chemical models. We use 
an extended and homogeneous data set of Galactic PNe to study the metallicity gradients and the Galactic structure and evolution.
The most up-to-date abundances, distances (calibrated with Magellanic Cloud PNe), and other parameters have been 
employed, together with a novel homogeneous morphological classification, to characterize the different PN populations.
We confirm that morphological classes
have a strong correlation with PN Peimbert's Type, and also with their distribution on the Galactic landscape. We studied the $\alpha$-element
distribution within the Galactic disk, and found 
that the best selected disk population (i.e., excluding bulge and halo component), together with the most reliable PN distance 
scale yields to a radial oxygen gradient of $\Delta$log(O/H)/$\Delta$R$_{\rm G}$=-0.023$\pm$0.006 dex kpc$^{-1}$ for the whole
disk sample, and of $\Delta$log(O/H)/$\Delta$R$_{\rm G}$= -0.035$\pm$0.024,
-0.023$\pm$0.005, and -0.011$\pm$0.013  dex kpc$^{-1}$ respectively for Type I, II, and III PNe, i.e., for high-, intermediate-, and low-mass
progenitors. Neon gradients for the same PN types confirm the trend. Accurate statistical analysis 
show moderately high uncertainties in the slopes, but also confirm the trend of steeper gradient for PNe with more massive 
progenitors, indicating a possible steepening with time of the Galactic disk metallicity gradient for what the $\alpha$-elements are concerned.
We found that the metallicity gradients are almost independent on the distance scale model used, as long as these scales are equally well 
calibrated with the Magellanic Clouds.  The PN metallicity gradients presented here are consistent with the local metallicity distribution; furthermore, oxygen gradients
determined with young and intermediate age PNe show good consistency with oxygen gradients derived respectively from other young (OB stars, H~II
regions) and intermediate (open cluster) Galactic populations.
We also extend the Galactic metallicity gradient comparison by 
revisiting the open cluster [Fe/H] data from high resolution spectroscopy. The analysis suggests that they could be compliant with
the same general picture of a steepening of gradient with time.

\end{abstract}


\section{Introduction}

It has been assumed over the years that the building of galactic disks is driven inside-out 
by the continuous accretion of in-falling gas (Larson 1976).  Evidences for the validity of the inside-out 
paradigm are still slim however, and are actively searched for. 
Inspection of detailed CDM simulations have shown that this picture may not be universal, 
and may vary from one galaxy to the other, to the point that some simulations show 
examples where galaxies can build their disk outside-in, at least partly (see e.g Sommer-Larsen et al. 2003, Roberston et al. 2004).  
In this respect, radial variations of properties of the Galactic disk may provide important observational
constraints, but this has been hampered by the lack of 
tracers for which distance can be measured with the relevant accuracy, explaining why the magnitude of 
abundance gradients in the disk is still actively debated. 

Nonetheless, it can be considered that a consensus on the existence of gradients
has been reached, and there is little doubt that inner regions are, in the mean, more metal rich than
the outer parts of the disk (Davies et al. 2009), and this effect can be seen both in the iron-peak and the $\alpha$-elements.
Also, recent understanding of the local metallicity distribution and kinematics strongly suggest 
that the solar radius is polluted by wanderers coming from the outer and inner disk, indicating
a rather strong radial variation of the metallicity (Haywood 2008).
Because radial metallicity gradients in galactic disks may be produced by a number of different 
processes, the measurement of how these gradients have evolved with time should provide an even
stronger constraint (e.~g., Fu et al. 2009), thus the importance of studying metallicity distributions of 
tracers of different Galactic ages.

Planetary nebulae (PNe) represent the Galactic stellar population with progenitors of turnoff mass M$_{\rm to}$ between $\sim$1 and 8 M$_{\odot}$, probing 
Galactic ages between $\sim3\times10^7$ yr  and $\geq$10 Gyr (Maraston 1998).  Planetary nebulae are distributed in the Galactic disk, bulge, and halo. Their Galactic distribution and
radial velocities offer
a first subdivision into populations, which can be studied separately. Furthermore, markers such as nitrogen abundances and morphological types allow to further discriminate  between disk PNe of relatively young and old progenitors. It is then sensible to undergo a study of Galactic metallicity for the different PN populations. This has been done in the past. Galactic disk PNe have been studied, among others,  by Kingsburg \& Barlow (1994) and, more recently, by Perinotto et al. (2004, P04), while bulge PNe are the subject of a focused study by Exter et al. (2004). The Galactic disk PNe have been the subject of many studies on metallicity gradients: $\alpha$-element abundances of PNe trace the
original progenitor composition, thus the chemical evolution of the Galactic disk. Perinotto \& Morbidelli (2006, PM06) reviewed all Galactic metallicity gradients published since the
seventies, and found that PNe in the Galactic disk trace a one dimensional, negative oxygen gradient  $\Delta$log(O/H)/$\Delta$R$_{\rm G}$=-0.07 (Faundez-Abans \& Maciel 1987) and -0.03 (Pasquali \& Perinotto 1993) dex kpc$^{-1}$. The excellent data revision by Perinotto and collaborators provides its own gradient of -0.016 dex kpc$^{-1}$ (PM06), which is shallower than
previously known. Almost contemporary Stanghellini et al. (2006) found a shallow gradient of -0.01 dex kpc$^{-1}$ for the oxygen abundances of Galactic disk PNe,  consistent with
PM06's analysis. 

The abundance analysis that yield to the gradients published so far might be marginally different, but hardly enough to make the metallicity gradients to differ outside the 
data uncertainties. A stronger factor for divergence is certainly the method of Galactic PN distance calculation, although PM06 showed that the gradients should result rather flat
independent on the distance scale used. The distance scale of PNe used is also a determining factor in the study of the different Galactic PN populations, their distance from the Galactic plane and their belonging to the thick disk, the bulge, the halo. Until a couple of years ago, distances to Galactic PNe where estimated by means of statistical distance scales, calibrated with
a few known distances to Galactic PNe. Now, with the availability of a wealth of spatially resolved images of Magellanic Cloud PNe from {\it HST}, the calibration has been done in a much
sounder way (Stanghellini et al. 2008, SSV). There is then purpose to reanalyze both the Galactic distribution of PNe and the metallicity gradients in a homogeneous way, with the distance scales based on the physical parameters of Magellanic Cloud PNe, whose distances are known.

In this paper we use the Magellanic Cloud-calibrated Galactic PN distance scale from SSV to examine the population of PNe in our Galaxy. We extend the P04 abundance database with all the abundances published more recently, and calculate the metallicity gradients for the different populations, and  homogeneously classify the Galactic PN morphology for further insight on population selection. In $\S$2 we describe the database built for this study; in $\S$3 we discuss the Galactic structure based on PNe; 
in $\S$4 we present the metallicity gradients from our analysis; sections 5 and 6 present respectively a discussion of our results and the conclusions of this work.

\section{The database}
\subsection{Sample selection and basic parameters} 

In order to build a database as complete as possible we started from Acker et al.'s (1994) catalog of Galactic PNe and include all confirmed PN therein. 
The catalog offers the Galactic coordinates and basic data for most nebulae. The nebular optical diameters, essential for determining the distances, have been
taken from Cahn et al.'s (1992, CKS) study, which takes into account the acquisition aperture, and, if not available therein, from Acker et al. (1994).
Furthermore, for extended Northern Galactic PNe the diameters have been taken from Manchado et al. (1996). The nebular flux at 5 GHz has been taken 
again from CKS preferentially, otherwise from Acker et al. (1994). If not available, the 5 GHz flux has been approximated by its relation to the H$\beta$ flux (see CKS).

\subsection{Distances and morphology}
As mention in the introduction the major reason for the reanalysis of the metallicity gradients is the advent of the new distance scale by SSV, which has been calibrated on the Magellanic Cloud PNe. In this paper we preferentially use the best, newly calibrated distance scale, which is the re-calibration of the CKS scale (Eq. 8a and 8b in SSV), but we also compare to other scales that use the ionized mass, the brightness temperature, and the surface brightness relations to the nebular radii (Eq. 5, 6, and 7 in SSV) to study the impact of different (but equally well-calibrated) scales on the metallicity gradients and the Galactic distributions. It is
worth noting that the choice of Magellanic-Cloud calibrated distance scales is based on the fact that these calibrations give the best results when compared to 
known individual PN distances, and they represent an improvement with respect to the old Galactic-calibrated scales (see Table 3 in SSV).

Planetary nebula morphology has been used successfully as a second parameter to determine the population type (Schwarz et al. 1993, Stanghellini et al. 1993, Stanghellini et al. 2002). The morphological classification available so far does not extend homogeneously to the complete sample of PNe used in the present study. We have then classified all PNe
homogeneously, based on the available [O III] $\lambda$5007 images available from Manchado et al. (1996), Schwarz et al. (1992), and other images all 
included in the online image collection by B. Balick\footnote{http://www.astro.washington.edu/users/balick/PNIC/}.
Knowing the morphological class allow us to exclude bipolar PNe from the sample used for gradient determination. In fact, SSV showed that bipolar PNe do not follow
the CKS calibration, and prefer to use the surface brightness- radius distance scale model for these PNe. On the other hand, it is very hard to define a radius for
bipolar PNe, thus their distance is always very uncertain, and it is just safer to exclude the extremely asymmetric PNe from the gradient analysis.

In Table 1 (published electronically) we give in column (1)  the PN names as in Acker et al. (1994); in column (2) the heliocentric distance calculated from the Magellanic Cloud  re-calibration of the CKS scale, except for bipolar PNe where we used the Magellanic Cloud re-calibration of the SB-R$_{\rm PN}$ relation, and where the uncertainties are assumed to be 20$\%$ for all PNe; column (3) gives the galactocentric distances (see below), column (4) the angular radii, column (5) the 5GHz flux, and column (6) the new morphological classification in major classes:
Round (R), Elliptical (E), Bipolar Core (BC), and Bipolar (B). In some cases the morphology was not defined, due to lack of clarity in the images or to very small nebular size. 

Galactocentric distances, R$_{\rm G}$,  are calculated by de-projecting heliocentric distances onto the Galactic plane with the usual relation
${\rm R_G=(R_{\odot}^2+(d~cos~b)^2-2R_{\odot}d~cos~b~cos~l)^{0.5}} $, where d is the PN heliocentric distance, R$_{\odot}$=8.0 kpc is the solar distance to the Galactic center, and b and l are the Galactic coordinates of the nebulae. While the statistical distance scale can not provide errors, SSV showed that uncertainties are well within 30$\%$ of the calibration. We assign a rather conservative uncertainty of $\pm$20$\%$ to the heliocentric distances, and we then propagate the error to the galactocentric distances, assuming  that the Galactic coordinates l and b have much lower relative uncertainties than the heliocentric distances. The galactocentric distances and their (formal) uncertainties are given in Table 1, column (3).

There are 1143 PNe in Acker's catalog, whose diameters are measured (either directly or as an upper limit) in 1053 cases. We were able to determine the
PN distances for $\sim$730 PNe.

\subsection{Abundances}

Perinotto et al. (2004) have carefully reviewed all elemental abundance analysis of Galactic PNe published since the seventies, and selected a sample that
is homogeneous in abundance determination methods and high quality of the observations (their sample A). We use the A sample from P04 in our study, augmented with the abundances published later than 
the P04 sample was analyzed, selecting the sets with the same criteria used by P04. Our updated sample includes abundances from (1) P04; (2) Stanghellini et al. (2006); (3) the series of papers by Pottasch  
and collaborators, summarized by Pottasch \& Bernard-Salas 2006; (4)  Kwitter \& Henry (2001), Milingo et al. (2002), and Kwitter et al. (2003); and (5) and Costa et al. (2004). 
It is worth noting that the series of papers by K. Kwitter and collaborators are 
formally dated before the P04 review, but actually the P04 work was completed before the former abundances have became available, thus not included in P04. 
There are several other papers on Galactic PN abundances published after 2004, but with the exception of the ones listed above they are reanalysis of earlier databases, thus
already included in P04, or excluded based on their selection criteria. Finally,  it is important to note that Stanghellini et al. (2006) and Pottasch \& Bernard-Salas (2006) abundances were calculated with the same method and ionization correction factors (ICF) than 
those of  P04, all utilizing the Kingsburg \& Barlow (1994) prescription. On the other hand, Kwitter \& Henry (2001), Milingo et al. (2002),  Kwitter et al. (2003), and Costa et al. (2004)
used a slightly a different ICF for oxygen. In this paper we correct for the differences in ICF(O) by using the ionic abundances given in the relevant papers. 
In the case when more than one reference give abundances for a particular PN we checked that the differences are within the uncertainties, and then chose the 
abundances preferring, after P04, all references whose abundances were originally calculated with the ICF relations by Kingsburg \& Barlow (1994).
We end up with a sample of 206 PNe with homogeneously calculated oxygen abundances, while the sample with neon abundances consists of 167 PNe. For more than half of the PNe of these sample we also have the helium and nitrogen abundance to determine whether belong to the Type I class (Peimbert \& Torres-Peimbert 1983).

We use the uncertainties in the abundances as quoted in the original papers, when available. 
Perinotto et al. (2004) gives formal errors for all abundances, which are used here directly. The abundances from Stanghellini et al. (2006) have uncertainties
of 0.04 dex for all elements, except for neon where the uncertainty is of the order of 0.1 dex.  Kwitter and collaborators assume that the typical uncertainty for the 
oxygen abundance is $\sim3\times10^{-4}$. In the cases where the uncertainties are not given (e.~g., Costa et al. 2004) we assume an uncertainty of 
0.15 dex for a conservative approach.

In Table 2 (published electronically) we list all PN abundances used in this paper. A PN is in this Table if at least one elemental abundances among the ones used
in this paper  (helium, nitrogen, oxygen, and neon) is available. Column (1) gives the PN name; Columns (2) through (5) give the elemental abundances and their uncertainties, derived as explained above, where He/H is given in linear form, while the other abundances are given in the usual 
form A(X)=log(X/H)+12.

\subsection{Populations}

The 1143 PNe in Acker's catalog belong to the thin disk, thick disk, halo, and bulge populations. In some of the applications below (i.e., disk metallicity gradients)
we select a pure disk population. We use the standard criteria to select bulge population PNe within angular distances less than 10 degrees from the Galactic center, angular radii smaller
than 10 arcsec, and 5 GHz fluxes smaller than 0.1 Jy (e.~g., Chiappini et al. 2009). There are 147 PNe in our  sample that satisfy all these conditions. 

Next we classify the PNe with determined abundances as Type I (with log(N/O)$>$-0.3 and  He/H$\ge$.125), Type II (non-Type I PNe with peculiar radial velocity V$_{\rm p}<$60 km s$^{-1}$), and Type III (non-Type I PNe with V$_{\rm p}>$60 km s$^{-1}$) as defined in P04. 
Peculiar velocities have been derived from measured radial velocities by
Durand et al. (1998) and Galactic rotation model therein, assuming a constant rotation for PNe with R$_{\rm G} > 2$R$_{\odot}$. The Type III PNe with altitude on the Galactic plane larger than 800 pc are
likely to be halo PNe. 

In Table 2, column (6), we give the disk population designation (Type I, II, or III) for the PNe, where possible, based on the abundances
and kinematics. For non-disk PNe we indicate whether they belong to the bulge or to the halo.

\section{Galactic structure}

In Figure 1 we show the sample of PNe with known distances in the R$_{\rm G}$--z plane, where z is the altitude on the Galactic plane. We plot the different major morphological types: R (round), E (elliptical), BC (bipolar core), and B (bipolar) in different panels, with all distances calculated with the Magellanic Cloud calibrated CKS scale. Using the most 
discordant distance scale, for example the Magellanic  Cloud-calibrated M$_{\rm ion}$-R$_{\rm PN}$ scale, gives a very similar distribution for all morphological types, with the
exception of bipolar PNe, whose statistical distances are not very accurate anyway. From Fig. 1 we confirm that the bipolar sample is confined close to the Galactic plane 
(again the distance bias for this class should be taken into account) and  the bipolar core PNe are intermediately distributed between elliptical and bipolar, as found in previous analyses (Stanghellini et al. 1993, 2002).  

In Figure 2 we plot on the same plane all the PNe, and then those of Peimbert Type I, II, and III. This plot has distances calculated in the same way than in Fig. 1, and includes  bipolar PNe. 

In order to assess the type of distance calibration we plot in Figure 3 the R$_{\rm G}$--z locus of all PNe with distances calculated with the method above, and then with the
T$_{\rm b}$--R$_{\rm PN}$, M$_{\rm ion}$--R$_{\rm PN}$, and SB--R$_{\rm PN}$ relations, using in all cases the superior Magellanic Cloud calibrations (Eq. 5, 6, and 7 in SSV). 
These samples  of 728 PNe include bipolar PNe. 
We can see that the different (Magellanic Cloud-calibrated) scales give broad agreement in the PN Galactic distribution, but there are differences, in that
PNe are significantly more clustered around the Sun when using Magellanic Cloud-calibrated CKS distances.

\section{Chemical  gradients in the PN population}

\subsection{One-dimensional Oxygen and Neon gradients}

The variation of the radially-binned $\alpha$-elemental abundances in disk galaxies is generally indicated as the {\it metallicity gradient}, and it is a measure of the composition 
of the Galaxy at the time of formation of the population under study. From the observational viewpoint, the metallicity gradient depends on the galaxy type, the population utilized, the method of calculating the $\alpha$-element abundances, the distances of the probes from the Galactic center, and also on the uncertainties associated to all these variables. Planetary nebulae well encompass several generations of stars in galaxies, thus are essential probes of metallicity gradient: their $\alpha$-elemental abundances are relatively easily determined, and the samples are sizable. On the other hand, historically it has been difficult to reach an agreement on the value of the 1D gradient slope from Galactic PNe, due to two factors: the unreliability of the distance scale, and the selection of the PN population. It is in fact essential that the gradients do not depend on the distance scale, but also that the disk population selected for this calculation is as complete and homogeneous as possible.
The use of oxygen and neon seems to be adequate in Galactic studies, where oxygen production through the AGB is marginal, while in low-metallicity galaxies oxygen might not be the ideal choice, since oxygen might be carried at the LIMS surface by the third dredge-up (Karakas et al. 2002). 

In Figures 4 through 6 we show the oxygen and neon abundances, in the usual form of A(X)=log(X/H)+12, versus the galactocentric distances of the PNe for the three
PN Types. The samples have been 
selected against both bulge and halo PNe, following the prescriptions of $\S$2.4. Bipolar PNe have not been included in the gradient plots to minimize the distance 
uncertainties of the disk PNe. Note that the bump around the solar vicinity in Figure 5 is due to the historical selection of nearby
PNe as spectral targets. 
In Table 3 we give the abundance averages and gradient resulting from our analysis with, in column (1) the $\alpha$-element considered, 
in column (2) the PN type included in the statistics, in column (3) the number of PNe in the sample (in parenthesis the sample size excluding bipolar PNe, i.~e., the sample size 
used for gradients and plots), in column (4)
the average linear abundance with its dispersion, in column (5) the dispersion of the y-axis distributions, in dex, in column (6) the gradient intercepts and their
uncertainties, and in column (7) the gradient slopes (in dex kpc$^{-1}$) and their uncertainties.  All samples analyzed here exclude halo and bulge PNe. Abundance averages include bipolar PNe, gradients and dispersions along the y-axis exclude all bipolar PNe. 

The gradients intercepts and slopes have been obtained with accurate fit. The statistical distance scales, used when handling a sizable number of PNe, do not provide errors for the individual distances. The impossibility to determine the distance uncertainties, except with ad hoc assumptions, induced us to estimate the fits using the abundance errors,
and varying the relative distance errors from 0 up to a reasonable maximum distance uncertainty. 
To do so we have used the routines {\it fit} and {\it fitexy}, described  in Numerical Recipes by Press et al. (1992).  Increasing the relative errors in the heliocentric distances from 0 to $\sim$30$\%$ will not change significantly the results of this paper. 
In Figure 7 we show, as an example, the effect of changing the distance uncertainties on the slope of the oxygen gradients for Type I (squares) and 
Type II (circles) PNe. Note how small is the effect on the gradient slope for the large sample of Type II PNe, where also the uncertainties in the slopes are much smaller
(yet much larger than the effect the distance uncertainty would have). On the other hand, for Type I PNe, where the gradient relies on a few PNe, the change of
slope with augmented uncertainties in the distances has a notable effect, especially since this particular sample of PNe has very few objects at large galactocentric 
distances, especially since the bipolar PNe have been excluded. This plot help us to underline the importance of the sample size as well as its homogeneity. It is very important noting that for distance uncertainties between
0 and $\sim$30$\%$ the gradient slopes for Type I, Type II, and Type III PNe are always in a sequence, the former steeper than the latter, both for oxygen and neon, even if, due to the rather large uncertainties in the Type I PN gradient slopes, the slopes of Type I and Type II PNe could be compatible.

The uncertainties to the fitted slopes given in columns (7) of Table 3 have been obtained by minimizing the $\chi^2$: we have reiterated the fits by artificially enlarging the 
data errors until  the $\chi^2$ probability, q($\chi^2$), was larger than 0.8. Note that the actual intercept and slope do not vary with this reiteration, but their uncertainties do. 

The metallicity gradients for the whole PN samples are all moderately flat, seemingly excluding gradient flattening with time when compared to gradients from other young stellar 
populations.
It is worth studying the gradients of PNe as a group, indicative of an intermediate-to-old age population, and also fractioned in the different PN populations, marking different evolutionary times of our Galaxy. 
Planetary nebula progenitors have masses from $\sim$1 to $\sim$8 M$_{\odot}$.  The lower mass limit is constrained by the dynamical time vs. evolutionary time scale for PN evolution (Stanghellini \& Renzini 2000), and by observations of PNe in old stellar systems (Alloin et al. 1976), while the upper limit, rather uncertain, marks the upper mass limit for a star to develop a degenerate CO core, and stellar evolution models find this limit to vary according to model assumptions. 
Type I PNe are  nitrogen enriched with respect to the general population. From an observational viewpoint, the minimum progenitor mass of Type I PNe is of the order of $\sim$2
 (Peimbert\& Serrano 1980) to $\sim$3 (Kaler et al. 1990) M$_{\odot}$, while evolutionary models predictions again depend on model assumptions, such as
 extra-mixing or overshooting. We assume that Type I PNe have  M$_{\rm to}\geq$2 M$_{\odot}$, corresponding to an approximate progenitor age t$\leq$1Gyr 
 derived from the turnoff mass- age relation for low- and intermediate-mass stars (Maraston 1998). It is worth noting that a mass limit of 3 M$_{\odot}$, corresponding to progenitors
 of t$\leq$0.3 Gyr,  would not change the impact of our findings.

We then assume Type II, and Type III disk PNe have progenitor masses in the approximate ranges 
of  2$> {\rm M_{to}} >$1.2, and M$_{\rm to}\leq1.2$ M$_{\odot}$ respectively (P04), with  progenitors 
ages approximately of  1$<$t$<$5, and t$\geq$5 Gyr (Maraston 1998).

When we take into account the different PN populations, we see that gradients are much steeper for the young Type I 
than for the old Type III disk PNe, consistently for both the oxygen and  the neon data sets. This result goes in the opposite direction to the flattening with time, which Maciel et al. (2003) have 
proposed based on PNe. We discuss this important result further in Section 5.

\subsection{Impact of distance scale and sample selection on the gradients}

In the previous section we noted that the gradients derived for non-Type I PNe are typically not steeper than -0.03 dex kpc$^{-1}$, consistent with the most recent determinations
(e.~g., Stanghellini et al. 2006, P06). To check the stability of our results it is worth going through the exercise of calculating the oxygen gradients with different assumptions. 

First, if we include the  few halo PNe in our sample the gradients would not change noticeably, even if it would change the meaning of the sample. 

A different argument should be applied for bulge PNe, since it is hard to separate bulge and disk populations. 
If we do not exclude bulge PNe
from our sample, and leave all other selection as for the set of Table 3,  we obtain oxygen gradient slopes of  -0.025$\pm$0.005 dex kpc$^{-1}$ for the complete sample, compared with 
-0.023$\pm$0.005 dex kpc$^{-1}$ obtained if we exclude the bulge population. For this and all other PN samples, the effect of including bulge PNe is to increase the (negative) gradient slopes. This could have contributed in part to the steeper gradients slopes of some of the published metallicity gradients from PNe, where an accurate selection of PN populations was not in effect.

Past metallicity gradient analyses have included bipolar PNe.
By including bipolar PNe in the gradient calculation we would obtain different gradients only for the Type I PNe
 ($\Delta$log(O/H)/$\Delta{\rm R_G}$=-0.024$\pm$0.02  dex kpc$^{-1}$), where the bipolar constitute a very large fraction of the population. 
Given the large uncertainties in distances to bipolar PNe, including them in even Type I PN gradients might be of limited significance. Including bipolar PNe in the sample would not change the global gradient nor that of Type II PNe in a significant way. 

Let us analyze what happens if we change the distance scale used to determine the metallicity gradients. We run the fit routines for all disk PNe and found:

 (a) $\rm {\Delta~log(O/H)/\Delta~R_G=-0.020\pm0.007}$ dex kpc$^{-1}$ by using the ${\rm Tb- R_{PN}}$ relation;
 
 (b)   ${\rm \Delta~log(O/H)/ \Delta~R_G=-0.020\pm0.008}$ dex kpc$^{-1}$ for the the ${\rm M_{ion}-R_{PN}}$ relation;
 
 (c)$ {\rm \Delta~log(O/H) / \Delta~R_G=-0.022\pm 0.008}$  dex kpc$^{-1}$ for the SB-R$_{\rm PN}$ relation, all  calibrated with Magellanic Cloud PNe (SSV).
 
We conclude that including bulge PNe steepens the metallicity gradient; including bipolar PNe has an impact but only for Type I PNe;  different statistical distance scale 
have a moderate impact on the gradients, if these scales are all calibrated on Magellanic Cloud PNe; and most importantly, the size of the statistical sample used to derive gradients 
might have an
impact on the slopes.

\subsection{The temporal evolution of $\alpha$-element gradients}

In Figure \ref{gradAge} we show the oxygen (left panel) and neon (right panel) metallicity gradient slopes against population age for the Type I, Type II, and Type III PN progenitors, for pure Galactic disk populations, bipolar excluded. To build this schematic plot we used the average PN population ages for the three Types, as described in $\S$4.1, where the bars represent the age span of each population, and the
gradient slopes with their uncertainties as reported in Table 3. All filled circles represent PN gradients from this paper. Note that the top x-axis gives the turnoff masses 
corresponding to the progenitor ages in the lower x-axis. 
The results clearly indicates that the magnitude of the gradient slopes of both oxygen and neon increase going from the older population PNe toward the 
young ones, indicating a steepening of the $\alpha$-element gradient with time in the disk of the Milky Way. The age intervals are not sharply defined, 
given the large uncertainties of data and models defining the different PN Types, especially for the young PNe, as discussed earlier in the paper. Nonetheless, given the scaling between ages and masses, choosing a lower limiting mass of 3 or 4 even M$_{\odot}$ instead of 2 M$_{\odot}$ for Type I PNe would not change the implications of Fig. 8
at all.
It is also worth noting that the dispersion in the log(O/H) along the Galactic disk varies with PN type, being smaller for Type II PNe and larger both for Types I and III (see Table 3). 

The gradient evolution with time relies on the results shown in Figures 4 through 6. While the Type II PN sample is well represented and unambiguously 
leads to the calculated gradient slope, the other PN Types should be discussed further.
First, the Type I PN gradient indeed seems to rely on a very few points, especially
at high galactocentric distances, with a single PN at 16 kpc skewing the slope. On the other hand, even excluding this Type I PN from the gradient calculation
we still obtain a steep (negative) slope (although the dispersion is higher, $ {\rm \Delta~log(O/H) / \Delta~R_G=-0.046\pm0.041}$  dex kpc$^{-1}$) confirming the steepening with time of gradients from Type II
to Type I PNe. 

Second, Type III PNe represent the very old disk population, thus the question of gradient blurring due to 
radial migration occurs. However, in Figure 6 we see that no PN seems to stand out as having a 
particularly high or low abundance for its location, 
which could be the case if migration had erased or flattened a significant gradient.
Moreover, comparison of Figures 4, 5 and 6 shows that most Type III PNe at R$_{\rm G}<$ 7kpc
have on average lower abundances than those seen at the same radii in the Type I and Type II populations.
We consider that it would require a particularly fine tuning of radial migration processes to 
confine low abundances Type III PN and remove higher abundances in this distance range, particularly
since such depletion is not seen for Type II population, which also must be affected by radial migration.

Figure 6 shows limited scatter (but significant dispersion);  we are aware that radial migration is an open issue, and it 
might be even higher than previously estimated (Ro{\v s}kar et al. 2008), thus 
it is not impossible that Type III PN gradient has been partially flattened
by stellar migration, but we consider it could hardly have erased a 
gradient that would have been higher than the one measured on Type II PN sample, in the case
the gradient evolution was one flattening with time.

\section{Discussion}

The PN metallicities and their gradients presented in this paper should be compared with 
abundances from other Galactic components. The overall comparison 
is limited for what the $\alpha$-elements are concerned, and it is presented in Section 5.2.

The existence of gradients should however also reflects on solar neighborhood stars, because
the local abundance distribution is not just the local sampling of stars born 
at solar galactocentric radius: although they dominate the local pool, the distribution 
is  a complex mix of stars born at different distances from the Galactic center. 
Somehow, the amplitude of measured gradients should be compatible with the range of abundances 
measured at the solar neighborhood. We address this question in Section 5.1.

\subsection{Consistency with the local metallicity distribution}

The spread of metallicity in the thin disk at the solar radius is mostly due to
non-evolutionary effects: the increase in metallicity due to chemical evolution cycling
over the last 8 Gyr (the age of the thin disk according to solar neighborhood data, see e.g Liu \& Chaboyer 2000) is limited to about [Fe/H]$\sim$0.4 
dex\footnote{The notation [X/H] is intended relative to the Sun} (from $\sim$ -0.2 to $\sim$0.2 dex), while the range of measured metallicity
is of the order of 1.2 dex (from $\sim$-0.6 to $\sim$0.6 dex).   How these extreme metallicities are compatible with the gradients measured in the previous
section?

The reason for this spread is to be found in stars that migrate across
the disk (Haywood 2008). Suggestions and results by Grenon (1972),  Sellwood \& Binney (2002), 
and Haywood (2008) conclusively point to the fact that local
metal-rich ([Fe/H]$>$0.2 dex) and metal-poor ([Fe/H]$<$-0.2 dex) 
stars of the thin disk have been brought to the solar radius by radial migration.
Sellwood \& Binney (2002)
proposed a mechanism that allows stars to individually exchange their angular 
momentum, allowing them to swap from one nearly circular orbit to another of different mean
orbital radius. At the solar radius, this mechanism involves a moderate fraction of stars, but it is important since it
gives a glimpse of the inner and outer Galactic disk chemistry.
The highest and lowest thin disk iron abundances measured in the solar neighborhood are respectively [Fe/H]$\sim$0.6 and $\sim$-0.6 dex, 
corresponding to [O/H]$\sim$0.3 and -0.3 dex (Ram\'irez et al. 2007,  Fig. 8), 
which roughly corresponds to the factor of 1/2 that is often used when scaling iron to oxygen abundances (e.g 
Martinelli \& Matteucci 2000, or more recently, Caimmi 2008).

The radial oxygen gradient for intermediate age PNe (Type II), $\sim$-0.023 dex kpc$^{-1}$,   would correspond
to [Fe/H]$\sim$-0.05  dex kpc$^{-1}$ if using the conversion factor illustrated above.  Given that the solar neighborhood mean metallicity is $\sim$0.0 dex,
such radial gradient  implies a mean iron abundance of about 0.25 dex at 5 kpc towards the inner disk (and -0.25 towards the outer disk). 
Allowing for an intrinsic dispersion similar to what is measured on the local iron distribution (0.15 dex), we easily reach (within 1 or 2 sigma) 
the highest metallicities that are observed at the solar radius, which means that the oxygen gradient measured on 
Type II PNe is well compatible with abundances measured in the local stellar population.

\subsection{Oxygen gradient: consistency with other disk populations}

In order to increase the significance of the PN results we explore the literature 
for object classes whose $\alpha$-element abundances and gradients have been studied. Several studies concerning
H II regions and young stars are available, while
the only possible comparison with an older population is that of open clusters. 

In Table 4
we list gradient slopes and their uncertainties (in column 2) and distance ranges (column 3) from different tracers (column 4), where
the authors have measured directly the oxygen abundances of the tracers.  Most tracers are very young stars, whose gradients have been 
plotted in Figure 8 with open circles. These young stellar data seem to fit in well with the young PNe of our sample, giving 
continuity to the plot. 

Our basic list of open clusters is the one designed by Magrini et al. (2009), to which we added a few more clusters from
the most recent literature. The various parameters are given in Table 5 with
columns (1) through (3) providing the name, age, distance from the Galactic center. Iron abundances,
referred to solar, are listed in column (4).  References to the iron abundances are as in Magrini et al. (2009). Columns (5) through (8) of Table 5
give the oxygen abundance ratios to iron, the oxygen abundances relative to solar, the actual oxygen abundance calculated for the solar
value given in the individual papers, and the reference for the oxygen abundances.

If we examine the clusters with nominal ages corresponding to the progenitors of the three PN Types, namely, in the age intervals of t$\leq$1, 1$<$t$<$5, and t$\geq$5 Gyr,
we can compare directly the PN results with those age-appropriate clusters whose oxygen abundances are available.  We have calculated 
the absolute oxygen abundances for the clusters by  using the same solar oxygen ratio used in the original papers. In cases where we
could not find the solar oxygen ratio used by the authors, 
we do not use the datum in the gradient determination, given that adopted values can differ by large amounts.

Oxygen abundances are available for a few very young clusters (t$\leq$1 Gyr) within a very limited galactocentric distance range, thus an estimate of the
gradient would not be meaningful (although the abundance distribution would be compatible, within the uncertainties, with that of Type I PNe).

A better comparison sample is available for the intermediate age clusters, corresponding approximately to the ages of Type II PNe progenitors, where the data span a 
range of galactocentric distances, and a gradient estimate is sensible.
We find that the oxygen gradient in open  clusters of ages between 1 and 5 Gyr is $\Delta$log(O/H)/$\Delta$R$_{\rm G}$=-0.028 dex kpc$^{-1}$. As discussed further in the next section, NGC 6253 and NGC 6791 can be suspected of having an origin in the inner disk, 
therefore not being truly representative of the metallicity at the radius where they are observed today.
Only NGC 6253 falls in the age interval considered here. If it is removed from the sample, the gradient decreases to -0.022  dex kpc$^{-1}$. The two estimates of 
metallicity gradient for open clusters of ages corresponding to Type II PN progenitors well encompass those derived directly from these PNe. As shown in Figure 8, oxygen
gradients from open clusters and PNe of intermediate ages are fully consistent with one another.  

Finally, for clusters older than 5 Gyr we have only three possible data points and limited galactocentric distance range, which do not 
lead to a gradient determination, although the calculated slope would be fully compatible with that of Type III PNe.

\subsection{Iron abundance in open clusters}

In the section above we have analyzed oxygen gradients in open clusters, and  
showed that a quantitative gradient  comparison with the PN population could be only achieved with the intermediate age populations, 
lacking the necessary data for direct comparison of PNe and open clusters in the young and old age bins.
For very young and very old clusters, we can
recur to a more qualitative analysis, introducing the iron abundances.

Figure 9 is a plot of  [Fe/H] vs.
galactocentric distances of the clusters in Table 5. Young clusters (t$\leq$1 Gyr) are
shown as squares, old ones (t$\geq$5 Gyr) as circles. Intermediate clusters are shown as diamonds. 
The very metal rich open
clusters NGC 6791 and NGC 6253 are specifically indicated.  The overall distribution has recently
been described as testifying a flattening of the metallicity gradient
in the outer disk (at R$_{\rm G}>$10 kpc), or even a plateau-like
distribution (Sestito et al.~2008 and references therein).  We
would like to propose another interpretation, suggested by the
behavior of the two ages groups, provided that another parameter is
taken into account, namely the possibility that some clusters, being on
eccentric orbits, are seen at a radius not representative of their origin 
(Magrini et al. 2009 offers a similar explanation). 
This is suggested in particular by the exotic nature of NGC 6791
and NGC 6253, the most metal-rich clusters near the solar
galactocentric radius (in Magrini et al. NGC 6253 is given an age of
4.3 Gyr, thus it is not included in the selection of old clusters, but its case is
very similar to NGC 6791).  NGC 6791 and
NGC 6253, with [Fe/H]  0.2 dex
above young metal-rich stars of the solar neighborhood, are possible
candidates for being intruders from the inner disk (see Bedin et al. 2006
for the possible origin of NGC 6791).  By inspecting Fig.~9 
we note that NGC 6791
is largely responsible for the impression that old clusters have a
steep gradient at R$_{\rm G}<$ 10-12 kpc. 
Once NGC 6791 is removed from the selection of old clusters, Fig. 9 illustrates that this 
population has a shallower gradient than usually invoked.
It is worth noting that cluster Collinder 261, with [Fe/H]=0.13 (Sestito et al. 2008),
could also be an intruder. However, its iron abundance is controversial (Friel et al. 2003 and 
Carretta et al. 2005 found [Fe/H] respectively -0.02 and -0.03 dex; we adopt the latter value in Figure 9).

The dichotomy between old and young clusters is then becoming more apparent, and suggests
that the gradient slope change noted at R$_{\rm G}\sim$10 kpc by several authors could result from the superposition of 
two different gradients for young and old clusters.
The distribution of intermediate age clusters shows  (expectedly) a combination of the two behaviors.

We have not tried to make a fit to quantify the gradients within the groups of clusters:
the number of objects in each group is small, and some of them are still being affected
by large errors, making it difficult to choose the correct values of metallicities and galactocentric
distances. Moreover, due to the lack of accurate information on the possible orbits of each cluster,
it would be difficult not to make an arbitrary selection of which clusters should be included or not 
in the old sample.

However, for illustrative purpose, we plotted different gradients on Figure 9 to see how
PN gradients are compatible with the above description.
The oxygen gradient derived
for PNe with progenitor younger than $\sim$1 Gyr is -0.035  dex kpc$^{-1}$. The old PN population, with 
progenitors older than $\sim$5 Gyr,  have flat metallicity gradient (-0.011 dex kpc$^{-1}$).
Allowing for the factor 2 that has been evoked when converting oxygen to iron abundances, we estimate an iron gradient
of the order of -0.07 dex kpc$^{-1}$ for the young population,  and a slope of $\sim$-0.02 dex kpc$^{-1}$ 
for the old population. These are plotted as continuous lines on Fig. 9.
Concerning the young population, we note good consistency between the gradients derived for Type I PNe and 
young clusters. This is also compatible with the gradient derived from cepheids, which amounts to -0.05 to -0.06 dex kpc$^{-1}$
(Lemasle et al. 2008, Luck et al. 2006). 
It is worth noting that the gradients for the old cluster population at the solar galactocentric 
radius reaches metallicities that correspond well to the metallicities of the oldest thin disk stars, 
at about $\sim$ -0.2 dex, according to the local age-metallicity relation (Haywood, 2006).

\subsection{Implications for Galactic evolution}

The PN metallicity gradients give a clear indication of gradients that steepen with time. 
Other stellar population sample that can be compared either directly - those with measured oxygen abundances - or indirectly
are limited.
They however lead to a qualitatively similar, if tentative, diagnostic: gradients apparently steepen
with time, at variance with previous suggestions (Maciel et al. 2003). While a sound comparison of our results with Maciel et al.'s
(2003) could not be performed until the PN database used by Maciel et al. is appropriately published,
there is purpose to explore the consequences of our findings and their implications for Galactic evolution.

We note that the change of slope, or the 'step' detected at R$_{\rm G}$=10 kpc in the metallicity distribution of
open clusters, 
which we interpret here as the superposition of a shallow gradient due to old clusters and a steeper one due 
to younger objects, has been detected in a number of other nearby galaxies.
This behavior has been presented in Vlaji{\'c} et al. (2009), as an argument 
for a flattening of gradients with time. Moreover,  Vlaji{\'c} et al. suggest that the metal-poor plateau in the outer disks could 
reflect the presence of an underlying old disk, formed at z$\sim$2. 
In the case of the Milky Way the clusters included in the selection at R$_{\rm G}>$ 10~kpc are mostly intermediate
age objects, having ages between 5 and 10 Gyr, thus they could not probe a disk formed at z$\sim$2. This does not
exclude the existence of such a primordial structure though: old open clusters could indeed originate from the slow evolution 
of a primordial gaseous disk.

The metallicity gradients calculated with PNe and other stellar populations should be used to constraint the chemical evolutionary models. These
models can reproduce both steepening and flattening of the radial metallicity gradient with time, depending of the specific
radial and time dependence of the star formation law and infall law, for which very little 
information is available, and sound data are essential to obtain insight on Galactic evolution.
Gradients of stellar metallicity in the Galactic disk may become steeper with time if infall occurs preferentially at larger
radii at later times, diluting metals in the outer disk (i.e., Tosi 2000). The problem with this scenario is that evidences
would be difficult to find, given the lack of possible direct observational clues about the temporal variation
of infall. 

Alternatively, a steepening of the gradients  with time could be understood if, starting from a similar metallicity, 
(1) the inner and outer parts of the Milky Way disk have evolved at different rates, or (2) the inner disk had more 
time to produce metals. In the latter case, we would expect the outer disk to be, on average, 
younger than the inner disk. This is difficult to check with the available data, but such a test could 
be within the reach of Gaia capabilities. A possible indication however may be found in the local stellar population, 
since radial migration of stars pollutes it with intruders from the outer disk. These stars are found 
in the same age range as stars with other metallicities (except for the youngest ones, which cannot be found locally because
of the time needed to migrate from the outer regions of the disk), as testified by the local age-metallicity distribution (Haywood, 2006).
In other words, there is no evidence that stars that have migrated from the outer disk cannot be as old as stars formed at the solar radii, 
which would be the case if the outer disk was significantly younger. 
In this respect, the thin disk star HD 203608, an example of star migrating from the outer disk, having a V-component 
space velocity in excess of the local standard of rest by 44 km s$^{-1}$, and a metallicity of -0.6 dex, has its age measured 
by asteroseismic constraints (Mosser et al. 2008) at 7.25 Gyr. 
Admittedly however, these facts are only suggestive, and cannot be taken as evidence that the outer disk is
as old as the inner thin disk, particularly since the time needed for significant migration in the disk 
introduces a bias favoring older stars and this is difficult to quantify.

The first scenario, where the inner and outer regions of the Galactic disk evolve at different rates, 
might be more appealing. The disk may have started to form stars at the same metallicity over the whole disk.
This is illustrated for example in the model of Sch{\"o}nrich \& Binney (2009).
When looking at their figure 6, a similar behavior is (qualitatively at least) observed: 
starting at the same initial metallicity, the disk evolves at different rates depending on the
distance to the Galactic center. The curves describing the increase of the metallicity at different radius
are first close to each other then slowly separate to reach increasingly different metallicity at later times.
Hence, a vertical cut through this plot would yield to steeper metallicity gradients for the young than the old stars.

The figure 6 of Sch{\"o}nrich \& Binney (2009) also shows that the initial increase (between ages 10-12 Gyr) in metallicity 
could be very steep, making it difficult observationally to distinguish whether the disk has started 
to form stars from a disk of gas with uniform metallicity as suggested by Vlaji{\'c} et al. (2009) or if a metallicity gradient 
pre-dated star formation. 

\section{Conclusions}

A study on PN metallicity gradients employing the most updated abundances and distance scale 
shows that gradients of $\alpha$-elements across the Galactic disk are moderate, and depend on the age of the PN population considered.
The amplitude of these gradients agree with the metallicity dispersion observed in the solar vicinity, which arises mainly
from the radial mixing of stars through the disk.
Planetary nebulae with young progenitors show steeper gradient slopes than in the old populations, indicating that $\alpha$-element gradients are steepening 
with time. The results are statistically sound, and do not depend very much on the distance scale, 
as long as it is Magellanic Cloud-calibrated.

Gradients  of oxygen  and  iron abundance  from  young (young stars and H~II
regions) and  intermediate-age  (open  clusters) populations agree with
those of PNe derived here, giving strength to the scenario of gradients
steepening with time. The oxygen gradient is found to increase from near-flat radial abundance distribution in Type III PNe, those whose progenitors
are in the lowest mass range for AGB stars, 
to -0.035 dex kpc$^{-1}$ for Type I PNe, those with the most massive progenitors.
These results contradict recent claim that gradients are flattening with time, and strongly suggest
that such constraints should be taken with some caution. 
The data on PNe are still sparse and the new distance scale, although significantly improved, could still be affected
by uncertainties. In addition, abundance gradients for older populations of PNe (in particular the Type III sample) might be affected by migration, which tends to flatten such gradients. However, we find it unlikely that gradients in the old PN population could reach
values as high as those found in some recent studies (e.g Maciel \& Costa 2009), 
and this is confirmed by the old open cluster population with high resolution spectroscopic abundance determinations.

\acknowledgments
Letizia Stanghellini warmly thanks Drs. Francoise Combes and Francois Hammer for their hospitality at the Observatoire de Paris (Paris and Meudon), 
where this work has been conceived and 
completed. We acknowledge scientific discussions with Drs. Bruce Balick, Laura Magrini, Karen Kwitter, Richard Henry, Angela Bragaglia, Monica Tosi, and Mark Dickinson.
We are grateful to an anonymous Referee for providing important comments to the earlier version of this paper.

\begin{figure}
\plotone{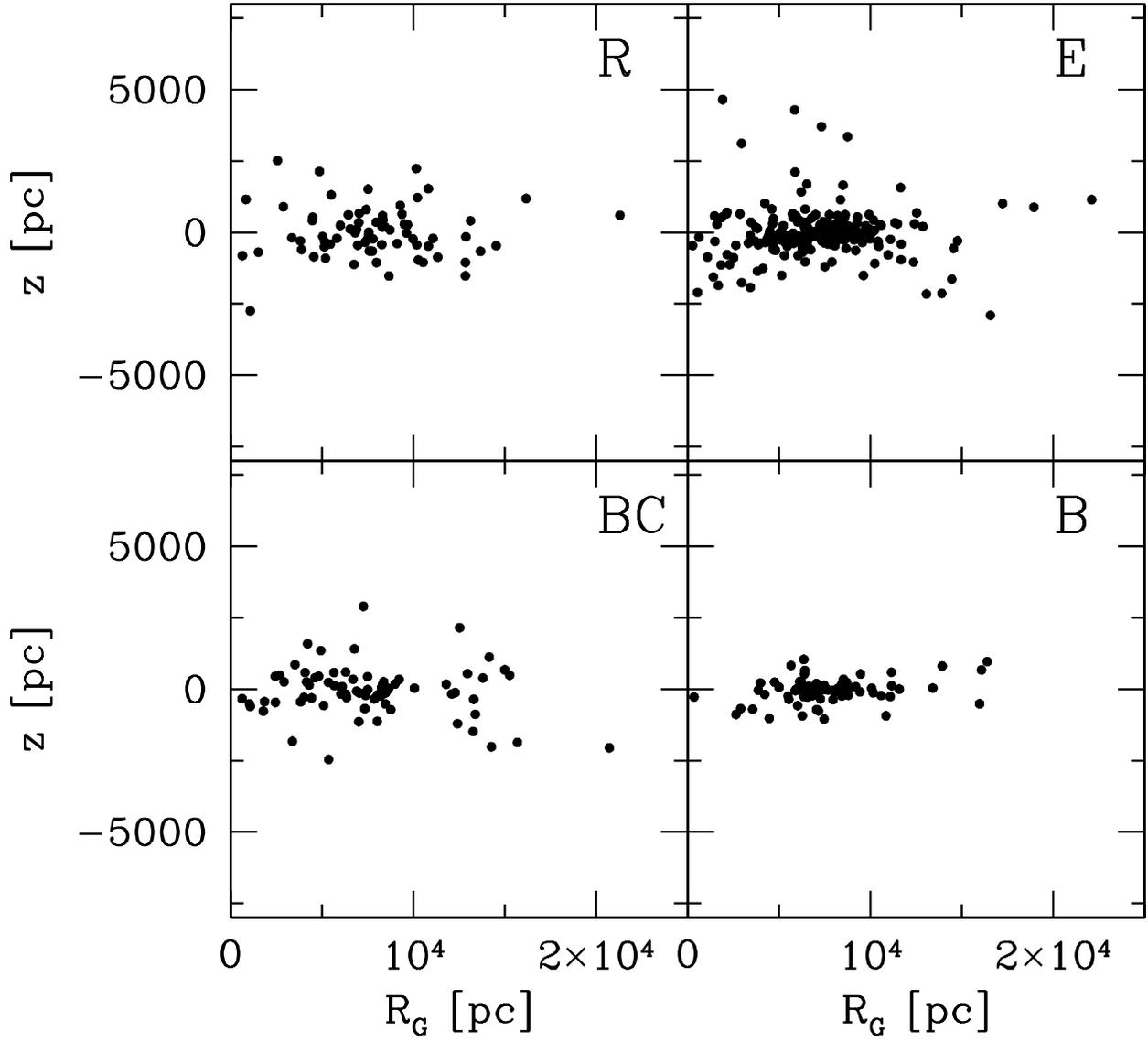}
\caption{Galactic distribution of morphological types in the R$_{\rm G}$--z plane. }
\label{rzmorph}
\end{figure}

\begin{figure}
\plotone{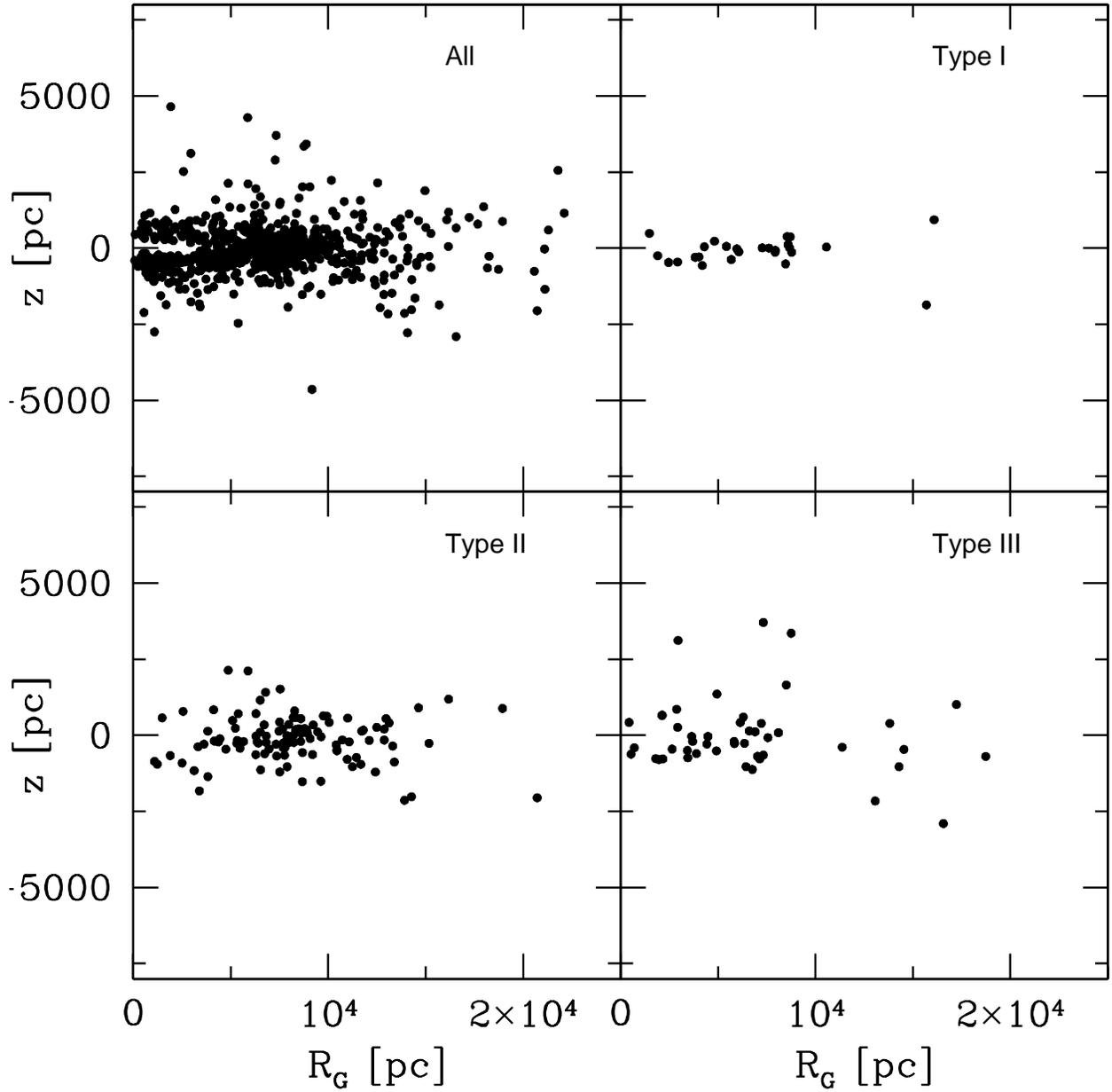}
\caption{Galactic distribution in the R$_{\rm G}$--z plane of all PNe with measured distances (top left), then Type I (top right), Type II (lower left), and Type III (lower right).  }
\label{rztype}
\end{figure}

\begin{figure}
\plotone{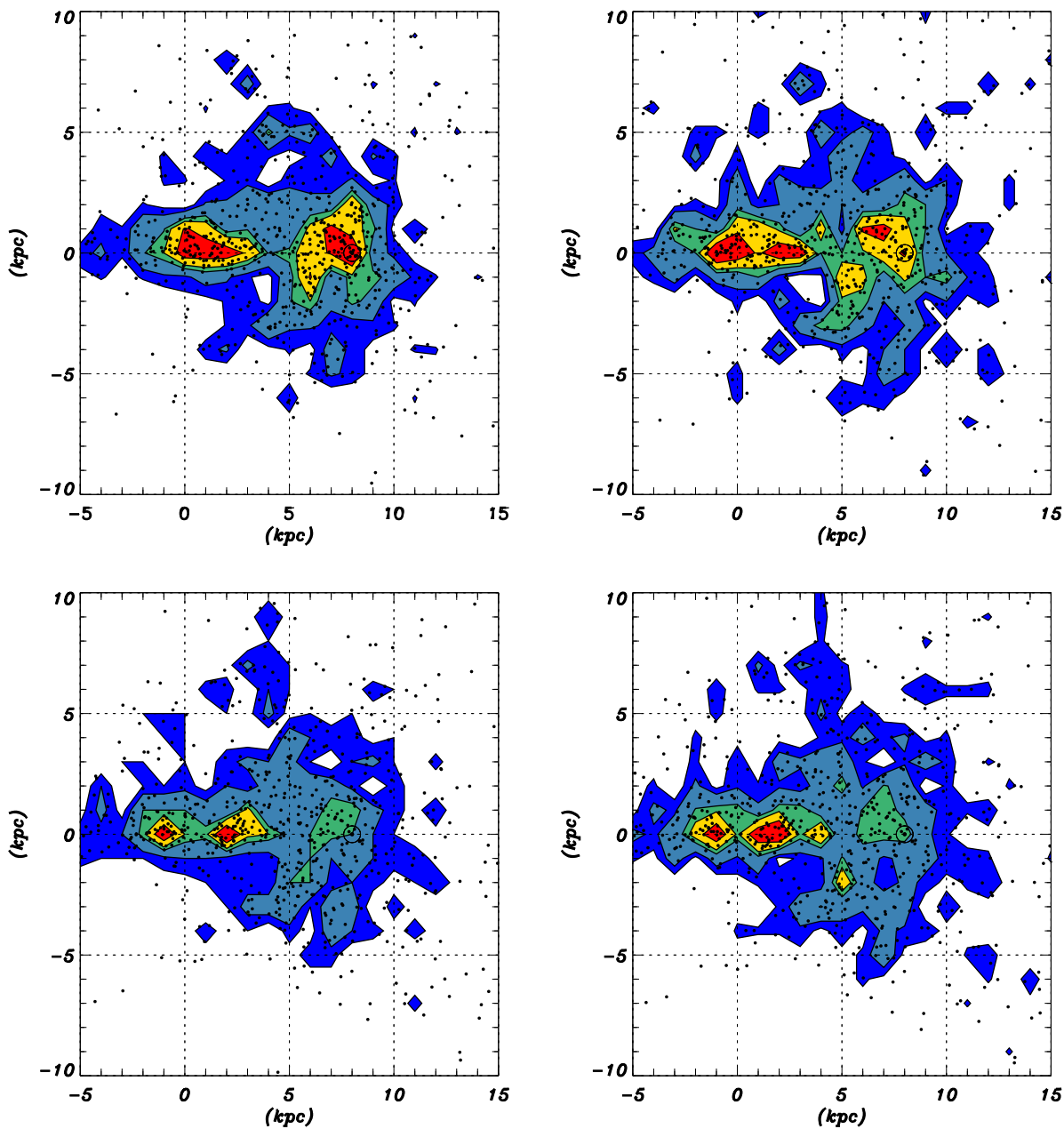}
\caption{Distribution in the Galactic plane for different distance scales. Contours give PN density, the Galactic center is at the origin. 
Each panel represent a different, Magellanic Cloud-calibrated, distance scale:
Upper left: CKS; upper right: SB-R$_{\rm PN}$; lower left: M$_{\rm ion}$-R$_{\rm PN}$; lower right: T$_{\rm b}$-R$_{\rm PN}$
scales.}
\label{xysurf}
\end{figure}

\begin{figure}
\plotone{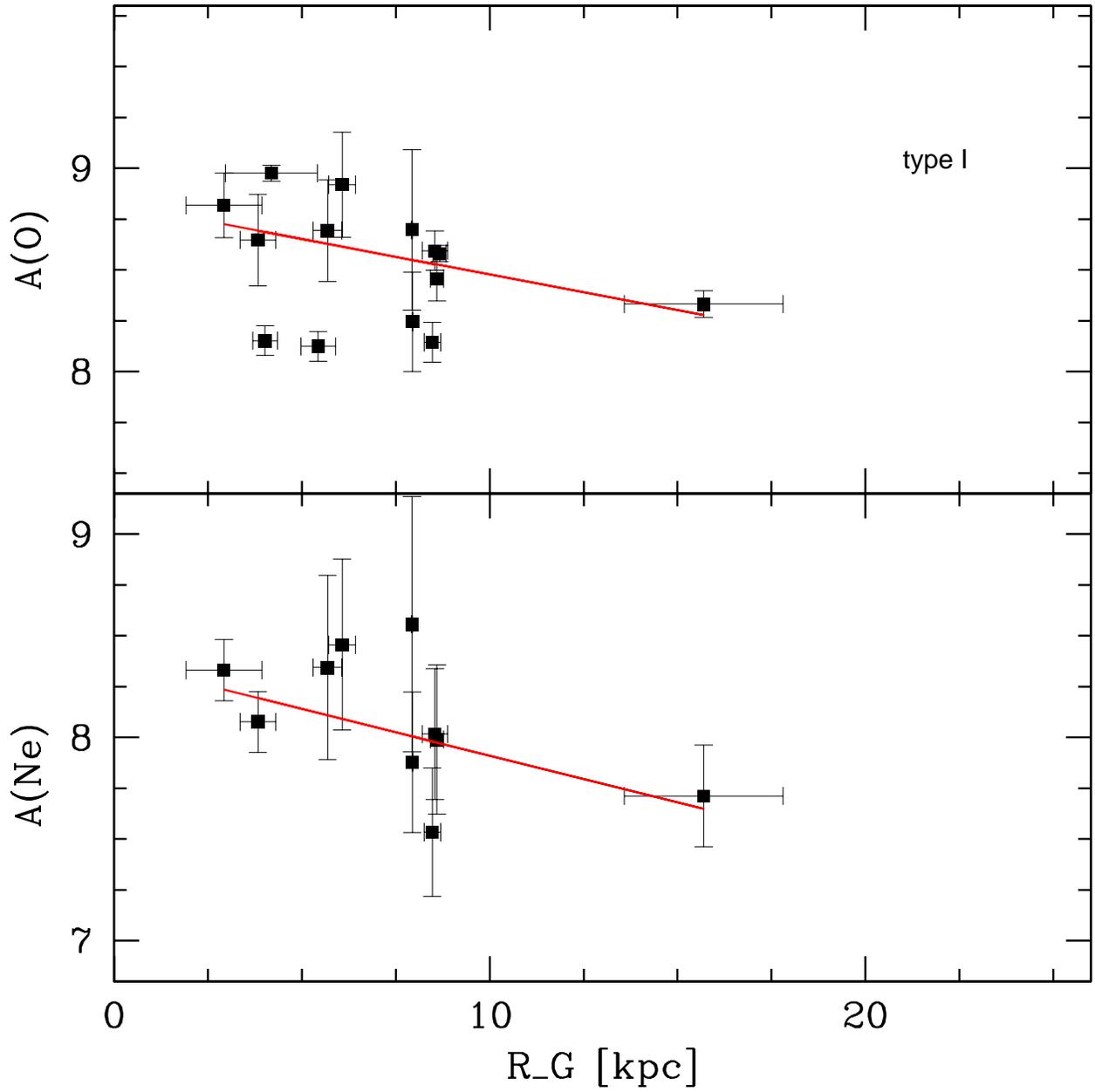}
\caption{The metallicity gradient of oxygen (top panel) and neon (bottom panel) for Type I PNe. Abundances (log(X/H)+12) vs. galactocentric
distances are plotted. This sample excludes bulge and halo PNe, and bipolar morphologies.
The solid line represents the least square fit and gives the slope reported in Table 3}
\label{1DgradI}
\end{figure}

\begin{figure}
\plotone{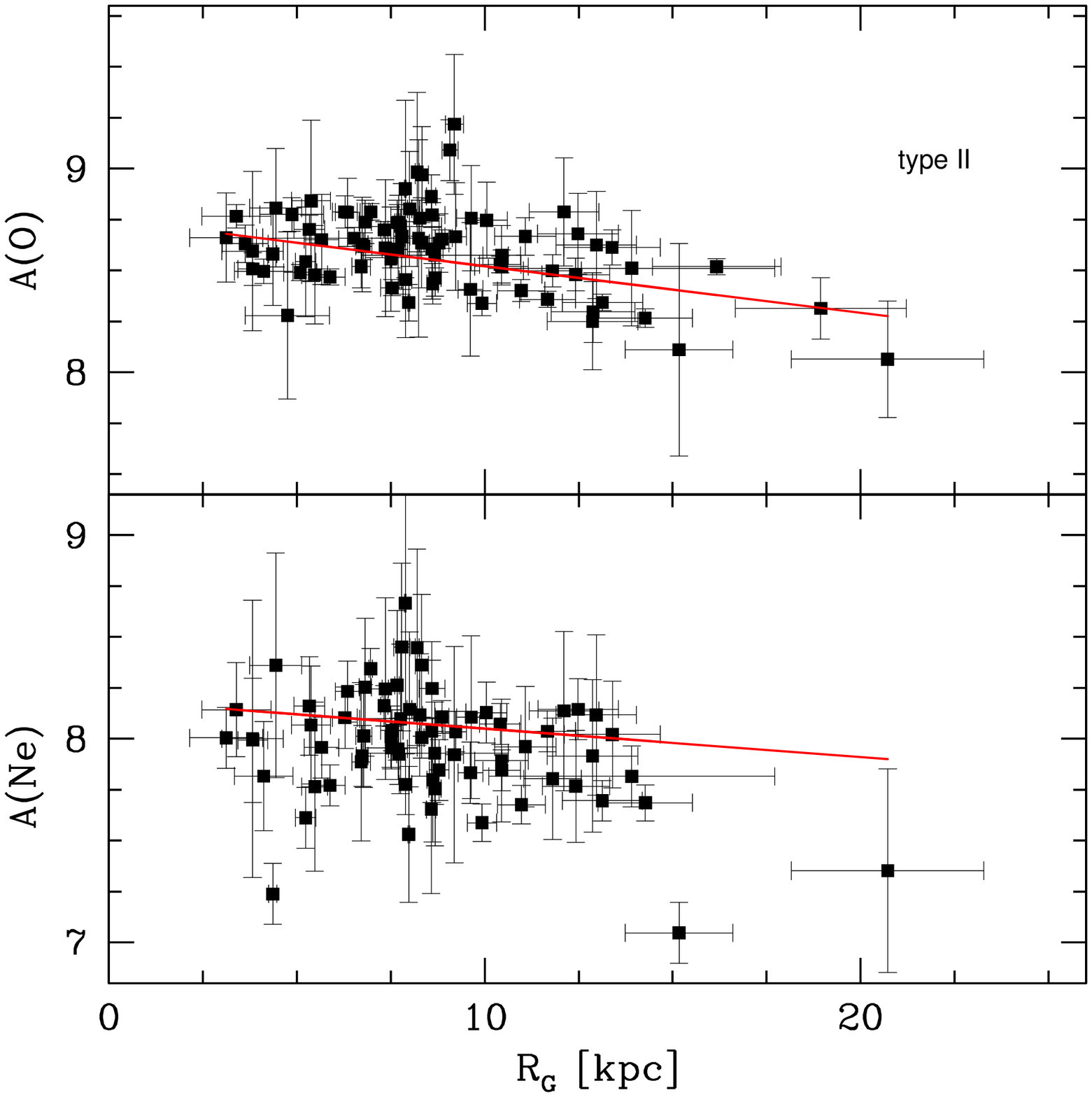}
\caption{As in Fig. 3, for Type II PNe. }
\label{1DgradII}
\end{figure}

\begin{figure}
\plotone{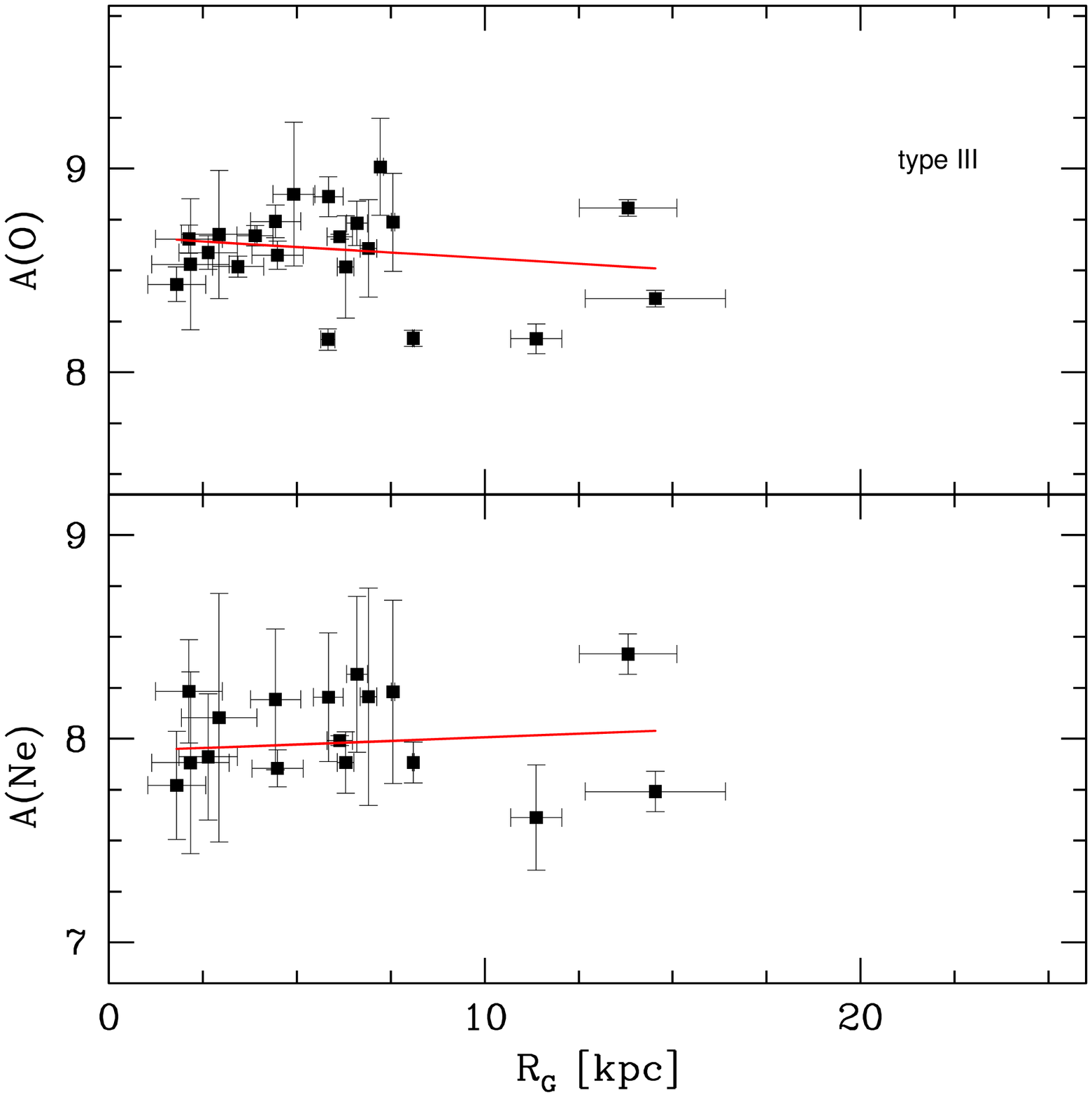}
\caption{As in Fig. 3, for Type III PNe}
\label{1DgradIII}
\end{figure}
 
\begin{figure}
\plotone{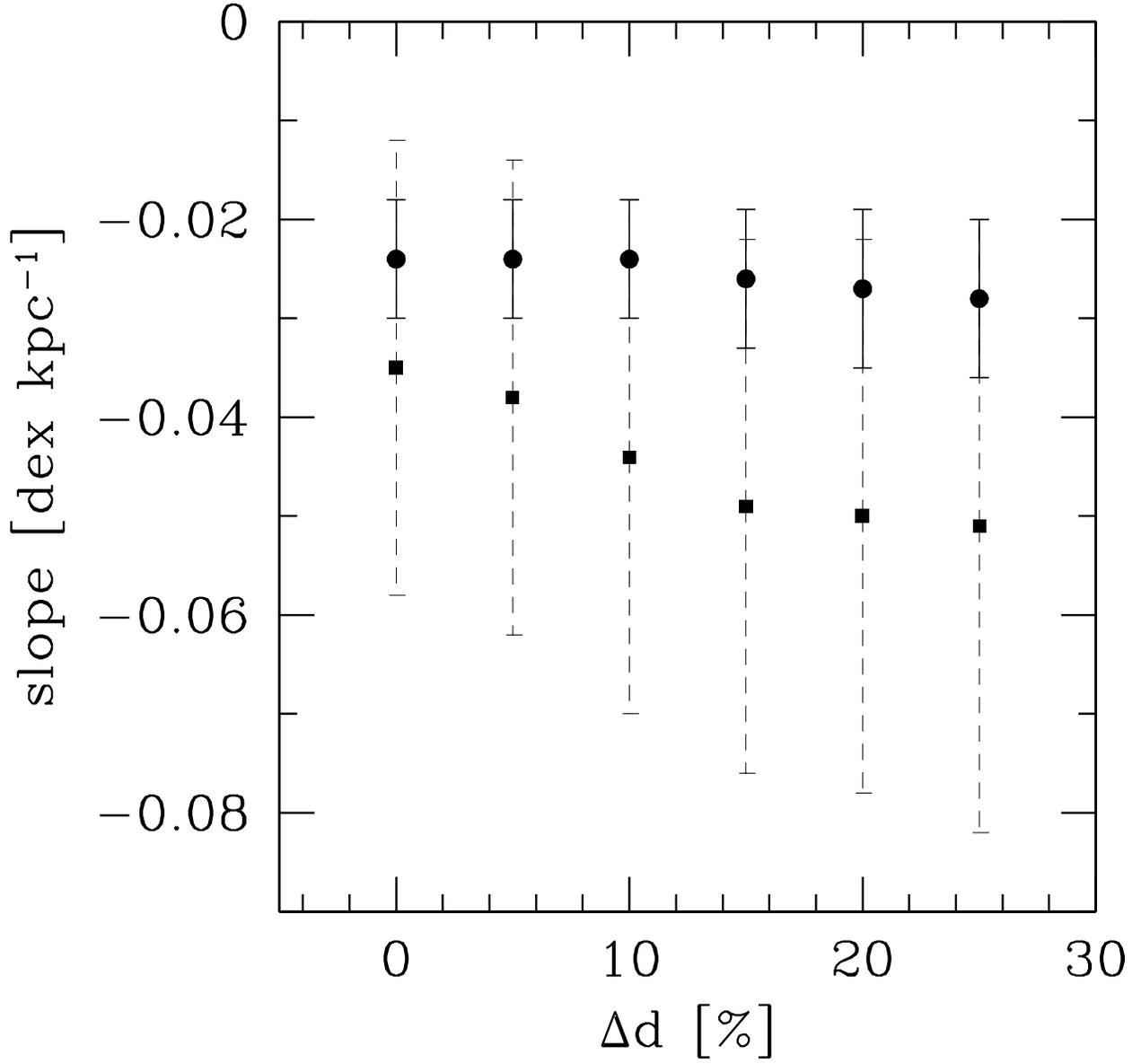}
\caption{Oxygen gradient slopes, in dex kpc$^{-1}$, vs. percentage error in the heliocentric PN distances. Type I PNe: squares, and dashed error-bars; Type II PNe: circles, and continuous error-bars. }
\label{fits}
\end{figure}

\begin{figure}
\plotone{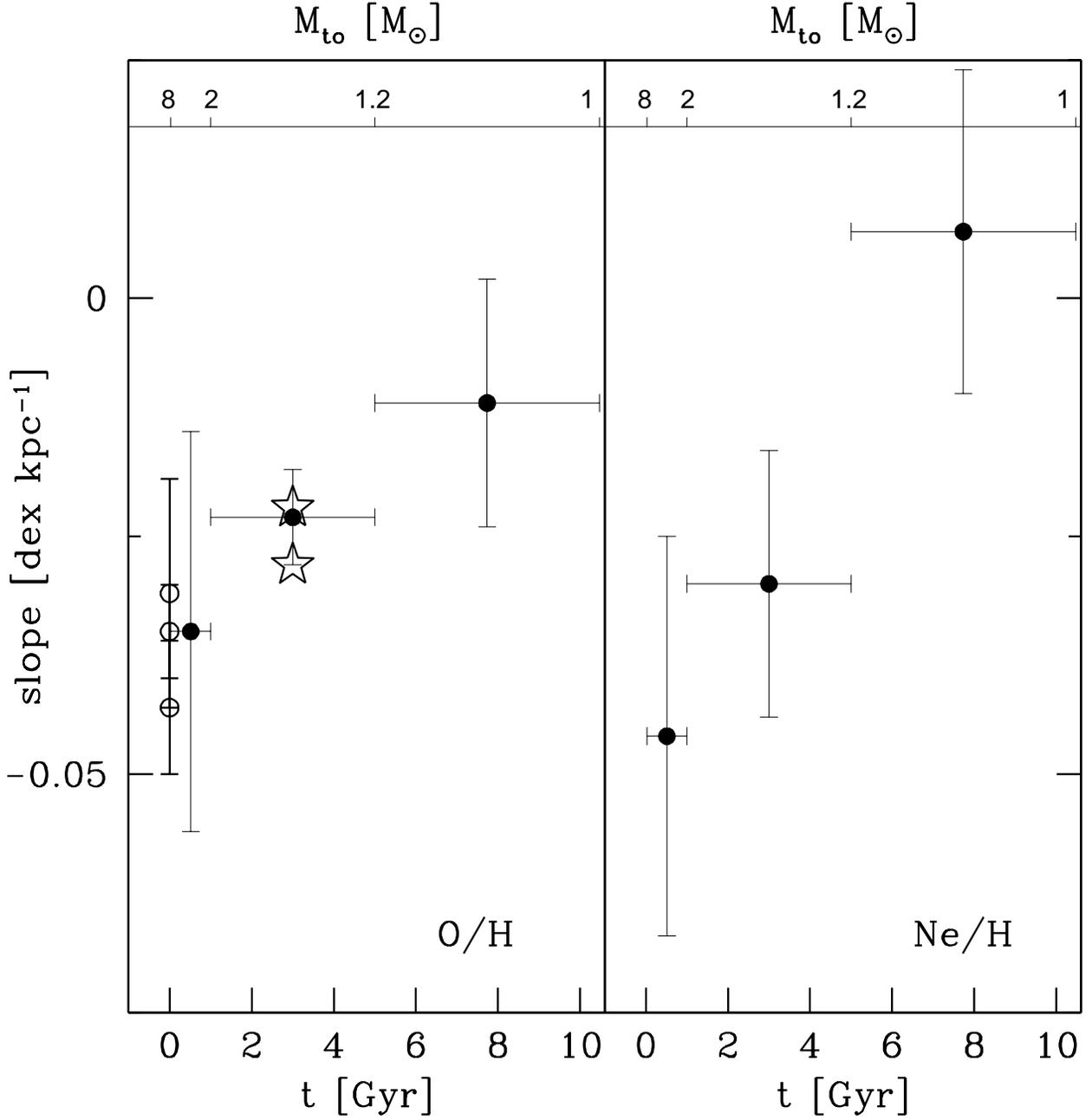}
\caption{The metallicity gradient slopes of oxygen (left panel) and neon (right panel) plotted against Type I, II, and III PN age ranges. 
Filled circles: PNe, this paper; open circles: young
stellar populations (see Table 4); stars: intermediate age open clusters, both including and excluding NGC 6253 from the sample (see Tables 4 and 5, and the text). 
The numbers near the top indicate the approximate progenitor turnoff mass, in M$_{\odot}$, corresponding to the x-axis ages. }
\label{gradAge}
\end{figure}

\begin{figure}
\plotone{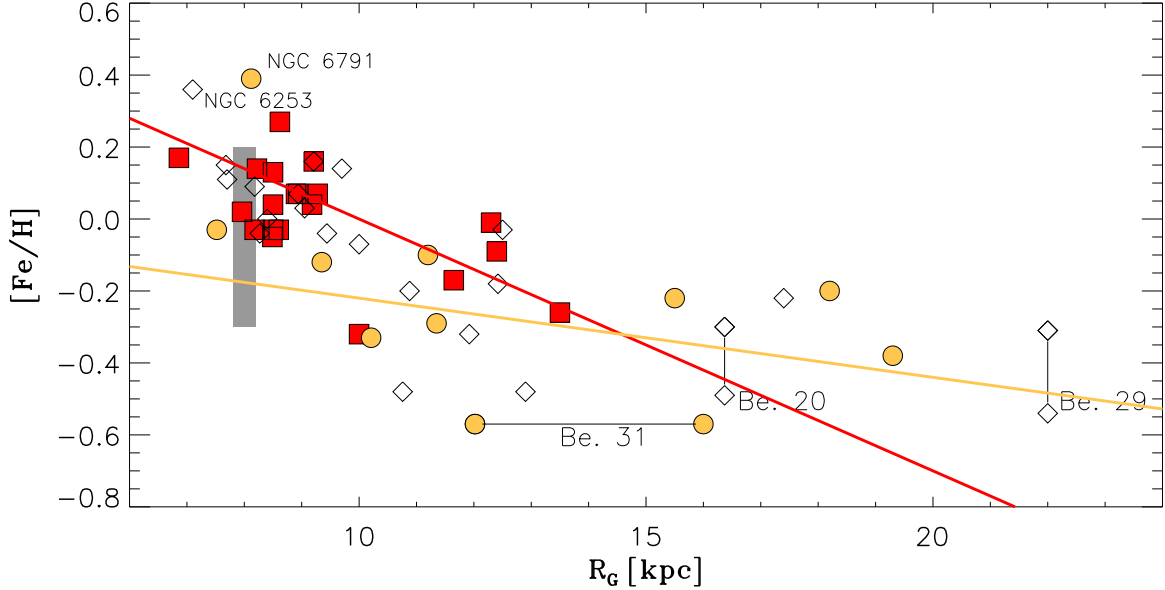}
\caption{Metallicity and galactocentric distance distribution for the open clusters of Table 5.
Squares are for clusters younger than 1 Gyr, filled circles for clusters older than 5 Gyr. Clusters with age
between these two limits are plotted as diamonds. Some clusters are given
two values of the metallicities or distances, illustrating significant uncertainties. Continuous lines
with slopes -0.07 and -0.02 show the metallicity gradients derived for Type I and Type III
PNe, scaled from oxygen.  The shaded box symbolizes the range 
of the metallicity evolution within the thin disk at the solar radius. The super metal-rich clusters NGC 6791 and 6253, and all clusters
with more than one abundance of distance determination, are specifically indicated.
}
\label{ocgrad}
\end{figure}

\clearpage

\begin{deluxetable}{cccccc}

\tablecaption{Distances, Radii, fluxes, and morphology}

\tablehead{ 
\colhead{PN~G} & \colhead{d}& \colhead{R$_{\rm G}$}& \colhead{$\theta$} & \colhead{F5GHz} &
\colhead{morph.}  \\
& [kpc]& [kpc]& [arcsec]& [Jy]& \\ 
(1)&(2)&(3)&(4)&(5)&(6)\\
}

\startdata
   003.9-14.9   &     8.2$\pm$1.6&       0.55$\pm$0.16&          2.0&        0.030&           E\\
     004.0-03.0    &    8.7$\pm$1.8&         0.93$\pm$0.14&         2.8&       0.008&         $\dots$\\
     004.0-05.8       & 6.6$\pm$1.3&        1.56$\pm$0.12&           6.0&      0.003&          $\dots$\\
     004.0-11.1       & 5.9$\pm$1.2&        2.30$\pm$1.20&         4.1&      0.018&           E\\

        \enddata
\end{deluxetable}

\begin{deluxetable}{cccccl}
\tablecaption{Abundances and PN type}

\tablehead{
\colhead{PN~G}& \colhead{He/H}& \colhead{log(N/H)+12}& \colhead{log(O/H)+12}&  \colhead{log(Ne/H)+12}& \colhead{PN Type}\\
(1)&(2)&(3)&(4)&(5)&(6)\\
}

\startdata
  000.3+12.2     &    10.99    $\pm$  0.013     &      7.68    $\pm$   0.356     &    8.50    $\pm$  0.059     &    7.82    $\pm$   0.268     &    Type II      \\
  002.0-13.4     &    11.04    $\pm$     0.150     &    8.18    $\pm$    0.150    &    8.66    $\pm$   0.220     &    8.01    $\pm$     0.150     &    Type II      \\
  002.2-02.7     &    10.97    $\pm$  0.013     &       7.99    $\pm$   0.373     &    8.39    $\pm$  0.038     &    7.70    $\pm$  0.073     &    bulge       \\

\enddata
\end{deluxetable}

\begin{deluxetable}{lrrrrrr}
\tablecaption{$\alpha$-element averages and radial metalicity gradients}
\tablehead{ 
\colhead{X/H}& \colhead{Type}& \colhead{N$_{\rm PN}$\tablenotemark{a}} & \colhead{${\rm <X/H>}$\tablenotemark{b}}&
\colhead{$\Delta$log(X/H)}& 
\colhead{log(X/H)$_0$}& \colhead{$\Delta {\rm log (X/H)}/\Delta {\rm R_G}$}\\
(1)& (2)&(3)&(4)&(5)&(6)&(7)\\
}

\startdata

O/H&  All& 		184  (145) &      4.24$\pm$2.21&   0.25&             8.71$\pm$0.06&     -0.023$\pm$0.006\\    						                      				

O/H& I&			23 (14)   & 	3.83$\pm$2.35&     0.26&               8.83$\pm$0.19&     -0.035$\pm$0.024\\     

O/H& II&              	103 (80)   &        4.49$\pm$2.28&	0.22&          8.75$\pm$0.05&     -0.023$\pm$0.005\\             
             
O/H& III&			29 (22)&        	4.39$\pm$2.19&	0.29&              8.67$\pm$0.09&     -0.011$\pm$0.013\\

Ne/H& All&		154 (120)&	1.14$\pm$0.76&	0.32&   7.99$\pm$0.09&     -0.012$\pm$0.011\\ 

Ne/H& I&			16 (10)&		1.50$\pm$0.90& 	0.28&  8.37$\pm$0.15&     -0.046$\pm$0.021  \\

Ne/H&  II&			91 (70) &         	1.16$\pm$0.78&	0.27&  8.19$\pm$0.13&     -0.03$\pm$0.014\\

Ne/H&  III&	        23 (17)&      	1.22$\pm$0.62&	 0.31&   7.93$\pm$0.12&      0.007$\pm$0.017\\

   \enddata

   \tablenotetext{a}{The sample size in parenthesis excludes bipolar PNe, which have been excluded in metallicity gradients.}
   \tablenotetext{b}{Average linear elemental abundances, multiplied by 10$^4$.}
\label{PNgradients}   
\end{deluxetable}

\begin{deluxetable}{cccc}
\tablecaption{Measurements of Oxygen radial gradients with different tracers}
\tablehead{ 
\colhead{Reference} &   \colhead{$\Delta {\rm log (O/H)}/\Delta {\rm R_G}$} & \colhead{dist. range}& \colhead{Tracer}  \\
                         & [dex kpc$^{-1}$]                    &      [kpc]        \\ 
                         (1)&(2)&(3)&(4)\\
}
\startdata
 Deharveng et al. (2000)  &  -0.035 $\pm$ 0.0049               &   5-15     &   H~II regions  \\

 Daflon \& Cunha (2004)   &  -0.031 $\pm$ 0.012        &   4.7-13.2 &   OB stars       \\
 Rood et al. (2007)       & -0.043 $\pm$ 0.007  &   0-17     &   H~II regions            \\

 This paper&  -0.028 to -0.022& 7-22& Open clusters, 1-5 Gyr \\
 \enddata
\end{deluxetable}

\clearpage
\begin{deluxetable}{lrrrrrrr}
\tablecaption{Open Clusters data}
\tablehead{ 
\colhead{Name}& \colhead{Age}&  \colhead{R$_{\rm G}$} & \colhead{[Fe/H]}&  \colhead{[O/Fe]}  &  \colhead{[O/H]} & \colhead{A(O)}  &  ref\tablenotemark{a}  \\
                    &        [Gyr]           & [kpc]           &            [dex]           &            [dex]       &   [dex]   \\
                    (1)&(2)&(3)&(4)&(5)&(6)&(7)&(8)\\ 
                    }
\startdata
IC 4665	     &   0.025  &  8.18  &   -0.03  &  0.35  &  0.32  & 9.19  & S05\\  
IC 2602	     &   0.035  &  8.49  &   -0.05  &                      &         & & \\
IC 2391	     &   0.053  &  8.50  &   -0.03  & & & & \\
Blanco 1     &   0.1    &  8.50  &   0.04  &   0.02    & 0.06  &  & Pa08\\
Pleiades     &   0.125	&  8.60  &   -0.03  &                          & -0.01 & 8.85 & S07 \\
NGC 6475     &   0.220	&  8.22  &   0.14  &  & &  &\\
M 11	     &   0.250	&  6.86  &   0.17  & & &   &\\
M 34	     &   0.250	&  8.90  &   0.07  & & 0.04  & 8.80  & Sc03\\
NGC 1883     &   0.65   & 12.3   &   -0.01   & -0.22  & -0.23 & 8.70 & J09 \\
NGC 3960     &   0.7    &  7.96  &   0.02  & & &  & \\
Hyades	     &   0.7    &  8.50  &   0.13 & -0.01&  0.12 & 8.81  & Sc06\\
Praesepe     &   0.7    &  8.62  &   0.27  &  -0.4&  -0.13&  8.53 &  Pa08\\
NGC 2324     &   0.7    & 11.65  &   -0.17  & & & & \\ 
NGC 2660     &   0.8    &  9.18  &   0.04  & & & & \\
Mel 71	     &   0.8    & 10.0   &   -0.32 &  -0.2&   -0.50 & 8.465 &  B96 \\
Ru 4	     &   0.8    & 12.40  &   -0.09  & & & &\\
Ru 7	     &   0.8    & 13.50  &   -0.26  & & &&\\
NGC 2360     &   0.9    &  9.28  &   0.07  & & & & H00 \\
\hline

NGC 2112     &	1.0     &  9.21  &   0.16  & & & &\\
NGC 2477     &	1.0     &  8.94  &   0.07  & & & & \\ 
NGC 7789     &	1.1     &  9.44  &   -0.04  &  -0.07 &   -0.11 & & T05 \\
NGC 6939     &  1.3     &  8.4   &   0.00   &  -0.11 & -0.11 & & \\
NGC 3680     &	1.4     &  8.27  &   -0.04  &  0.2 &   0.16 & 8.82 & Pa08 \\
NGC 6134     &	1.6     &  7.68  &   0.15  & & & &\\
IC 4651	     &   1.7    &  7.70  &   0.11  & & & &\\
NGC 1817     &   1.7    & 10.0   &   -0.07    & -0.19 & -0.25 & 8.68 & J09\\
Be 73	     &   1.9    & 17.40  &   -0.22  & & && \\
NGC 2158     &   1.9    & 12.5   &   -0.03  & -0.17  & -0.20 & 8.73 & J09\\
NGC 2506     &	2.2     & 10.88  &   -0.20  & & && \\
NGC 2141     &	2.5     & 12.42  &   -0.18  &   0.00 &   -0.18& 8.78 &  Y05\\
NGC 6819     &	2.9     &  8.18  &   0.09  & & & &\\
Be 66	     &   3.7    & 12.9   &   -0.48  & & & &\\
Be 22	     &   4.2    & 11.92  &   -0.32  & & & &\\
M 67	     &   4.3    &  9.05  &   0.03  &  0.01  &   0.04 & 8.70 & R06\\
Be 20	     &   4.3    & 16.37  &   -0.30; -0.49  &  0.18 &     -0.31 & 8.55 & S08; Y05 \\ 
Be 29	     &   4.3    & 22.0   &   -0.31; -0.54  &  0.23&    -0.31 & 8.55 & S08; Y05 \\ 
NGC 7142     &	4.4     &  9.70  &   0.14  &  -0.10&    0.04 & 8.97 & J08\\
NGC 6253     &	4.5     &  7.10  &   0.36  &  -0.18&    0.28& 9.07 & C07 \\
NGC 2243     &	4.7     & 10.76  &   -0.48  &  0.18&     -0.30 & & GC94\\
\hline
Be 75	     &   5.1    & 15.50  &   -0.22  & & & & \\
Be 31	     &   5.2    & 12-16 &   -0.57  &   0.24 &  -0.33& & Y05 \\ 
Mel 66	     &   5.4    & 10.21  &   -0.33  & & & & \\
Be 32	     &   6.1    & 11.35  &   -0.29  & & && Br08\\
Be 25	     &   6.1    & 18.20  &   -0.20  & & &&\\
NGC 188      &	6.3     &  9.35  &   -0.12  & & && Ho90\\
Saurer  1    &    6.6   & 19.3   &   -0.38  &  0.47  &   0.09 & & Pa08  \\
Cr 261	     &   8.4    &  7.52  &   -0.03; 0.13  & -0.12  &  -0.15 & 8.64 &  C05; S08  \\
Be 17	     &   10.0   & 11.2   &   -0.10  &  0.00    & -0.10     & 8.83 & F05 \\
NGC 6791     &	10.9    & 8.12   &    0.39  & -0.31     &   0.16   & 8.95 & C07\\

 \enddata
 
\tablenotetext{a}{
Reference codes: 
B96: Brown et al. (1996); 
Br08: Bragaglia et al. (2008);
C05: Carretta et al. (2005); 
C07: Carretta et al. (2007); 
F05 : Friel et al. (2005);
GC94: Gratton \& Contarini (1994); 
H00: Hamdani et al. (2000);
Ho90: Hobbs et al. (1990); 
J07: Jacobson et al. (2007); 
J08: Jacobson et al. (2008); 
J09: Jacobson et al. (2009); 
Pa08: Pace et al. (2008); 
R06: Randich et al. (2006); 
S05: Shen et al. (2005); 
S07: Shen et al. (2007);
Sc03: Schuler et al. (2003); 
Sc06: Schuler et al. (2006);  
T05:Tautvaisiene et al. 2005; 
Y05: Yong et al. (2005).
S08: Sestito et al. (2008)
}

\label{ocdata}
\end{deluxetable}

\end{document}